\documentclass[journal]{IEEEtran}

\usepackage{cite}
\usepackage{multicol,lipsum}
\usepackage{amsmath,amssymb,amsfonts}
\usepackage{algorithmic}
\usepackage{graphicx}
\usepackage{textcomp}
\usepackage{xcolor}
\usepackage{booktabs}
 \usepackage{hyperref}
\usepackage{enumerate}
\usepackage{algorithm}
\usepackage{multirow}
\usepackage{standalone}
\usepackage{tikz}
\usetikzlibrary{positioning}
\usetikzlibrary{decorations.markings}
\usepackage{float}

  \usepackage{enumerate}
 \usepackage{caption}
\usepackage{subcaption}
 \usepackage{setspace}
\usepackage{verbatim}
   \usepackage{tabularx}
 \usepackage{makecell}

 \usepackage{cite}
\usepackage{multicol,lipsum}
\usepackage{amsmath,amssymb,amsfonts}
\usepackage{xcolor}
\usepackage{booktabs}
 \usepackage{enumerate}
\usepackage{algorithm}
\usepackage{algorithmic}
\usepackage{multirow}
\usepackage{url}
\usepackage{hyperref}
\usepackage{standalone}
\usepackage{tikz}
\usetikzlibrary{positioning}
\usetikzlibrary{decorations.markings}
\usepackage{caption}
\usepackage{subcaption}
 \usepackage{makecell}
\usepackage{mathrsfs}  

 \usepackage{color}

\begin{document}
 
\title{Towards Standardizing Affine Frequency Division Multiplexing (AFDM) for
Future  Wireless Networks}

\author{Qu Luo,~\IEEEmembership{Member,~IEEE,} Lixia Xiao, ~\IEEEmembership{Member,~IEEE,} Pei Xiao,~\IEEEmembership{Senior Member,~IEEE,}    
Zilong Liu,~\IEEEmembership{Senior Member,~IEEE,} Yin Xu,~\IEEEmembership{Member,~IEEE,}   Qihao Peng,~\IEEEmembership{Member,~IEEE,}
Zeping Sui,~\IEEEmembership{Member,~IEEE,}
Hee Wook Kim,~\IEEEmembership{Member,~IEEE,}    and Hüseyin Arslan,~\IEEEmembership{Fellow,~IEEE}.
 
 \thanks{Qu Luo, Pei Xiao and Qihao Peng    are with 5G \& 6G Innovation Centre, University of Surrey, Guildford, GU2 7XH, UK. (e-mail: {q.u.luo, p.xiao, q.peng }@surrey.ac.uk).}
 \thanks{Lixia Xiao is with Research Center of 6G Mobile Communications, School of Cyber
Science and Engineering, Huazhong University of Science and Technology,
Wuhan 430074, China   (e-mail:lixiaxiao@hust.edu.cn.)}
\thanks{  Zilong Liu and Zeping Sui are with the School of Computer Science and Electronics Engineering, University of Essex, Colchester CO4 3SQ, U.K. (e-mail:  zilong.liu@essex.ac.uk,zepingsui@outlook.com).}
\thanks{ Yin Xu is with the School of Information Science and Electronic Engineering, Shanghai Jiao Tong University, Shanghai 200240, China, (e-mail: xuyin@sjtu.edu.cn).}

\thanks{Hee Wook Kim is with the Satellite Communication Research Division, Electronics and Telecommunications Research Institute (ETRI), Daejeon 34164, South Korea (email: prince304@etri.re.kr).}

\thanks{H. Arslan is with the Department of Electrical and
Electronics Engineering, Istanbul Medipol University, Istanbul, 34810, Turkey, (e-mail: huseyinarslan@medipol.edu.tr).}

}
\maketitle

\begin{abstract}
Affine frequency
division multiplexing~(AFDM) has   emerged as a compelling
waveform candidate for  future wireless networks, owing to its strong resilience to doubly
selective channels and its ability to enable the seamless integration
of communication and sensing functionalities.
Against this context, this  article provides a systematic study of AFDM from a
standardization perspective. We first introduce the principles of
AFDM and discuss the major considerations involved in waveform
standardization. We then examine the backwards compatibility of AFDM with 4G/5G multi-numerology frameworks and their anticipated evolution, frequency-modulated continuous-wave (FMCW) radar waveforms, and long-range (LoRa) modulation, demonstrating that AFDM can be incorporated into legacy processing chains with limited modification. Key standardization-critical capabilities
are further discussed, including multiple-antenna and multi-user
support, and peak-to-average power ratio~(PAPR). Finally, we
investigate the potential of AFDM in several  emerging
scenarios, including non-terrestrial networks~(NTN), integrated
sensing and communications~(ISAC), vehicle-to-everything~(V2X), and
underwater acoustic~(UWA) communications, whereby  severe delay-Doppler
dispersion places stringent demands on waveform robustness. Through these explorations, it is shown  that
  that AFDM represents a timely and
compelling technology for future wireless networks.
\end{abstract}

\begin{IEEEkeywords}
Affine frequency division multiplexing (AFDM), standardization, NR  multi-numerology,    FMCW, NTN, ISAC, V2X and UWA.
\end{IEEEkeywords}

\IEEEpeerreviewmaketitle

\section{Introduction}
\IEEEPARstart{F}{uture} wireless networks are expected to support an intelligent, highly interconnected society with tightly integrated communication, sensing, and computing across terrestrial and non-terrestrial platforms \cite{Luo2026ChirpBasedOCDMAFDM}. The IMT-2030 vision anticipates the sixth-generation (6G) system  delivering peak data rates beyond 1~Tbps, sub-millisecond latency, connection densities up to $10^8$ devices/km$^2$, and reliable operation at mobility speeds exceeding $1000$~km/h~\cite{ITU_R_M2160_2023}. A defining requirement is to    achieve ubiquitous connectivity   in challenging  high-mobility scenarios, including high-speed railways, unmanned aerial vehicles~(UAVs), autonomous driving platforms, and integrated terrestrial and non-terrestrial networks~(NTNs). Achieving these ambitious targets  demands fundamental innovation at the physical layer (PHY), with multicarrier waveform design playing a central role \cite{11164334,Sui2025MultiFunctionalChirp}. The widely deployed
orthogonal frequency division multiplexing (OFDM) systems
may be ineffective and unviable due to the significant intercarrier interference (ICI) \cite{PerofrmanceOFDM}. Besides, in high frequency bands (e.g., millimeter frequency or even THz bands), OFDM generally suffers from significantly higher phase noise and carrier frequency offset, leading to further increase of ICI. These limitations have motivated an active search for alternative waveforms and waveform enhancements that are inherently more resilient high mobility, channel dynamics, and hardware impairments.

 \begin{figure*}[htbp]
    \centering
    \includegraphics[width=1\linewidth]{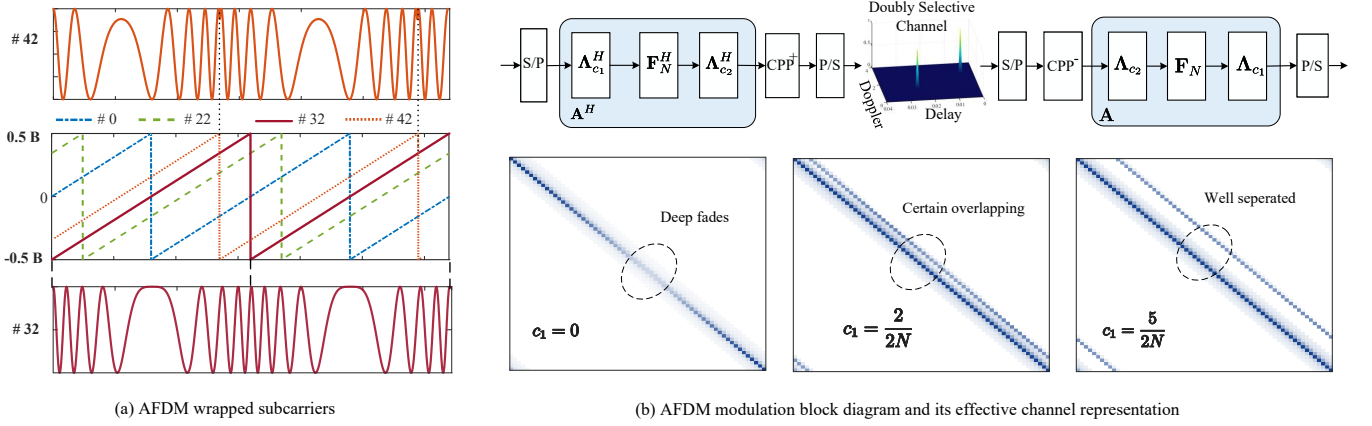}
    \caption{Illustration of  AFDM Principles: (The AFDM subcarriers with $N=64$ and $c_1 = \frac{1}{N}$); (b) AFDM modulation and its effective channel representation fore $N=64$, assuming  a two-path channel with normalized delay-Doppler pairs $(0, 0.5)$, $(2, 0.1)$, respectively. S/P and P/S denote serial-to-parallel and parallel-to-serial conversion, respectively. }
    \label{AFDMTro}
\end{figure*}

Among the emerging waveform candidates for highly dynamic wireless environments, affine frequency division multiplexing~(AFDM) has recently attracted significant attention~\cite{Luo2026ChirpBasedOCDMAFDM,bemani2023affine}. AFDM maps data symbols onto orthogonal chirp subcarriers through the inverse discrete affine Fourier transform~(IDAFT), and introduces two tunable chirp parameters, $c_1$ and $c_2$, to  provide additional flexibility in waveform design. With appropriate parameter configuration matched to the channel delay-Doppler spread, AFDM yields a quasi-static and separable effective channel representation in the DAFT domain, thus can achieve  full diversity over doubly dispersive channels. Moreover, AFDM generalizes OFDM and orthogonal chirp division multiplexing~(OCDM) as special cases, offering a unified waveform framework \cite{Luo2026ChirpBasedOCDMAFDM}. Its transceiver can be implemented by augmenting a conventional FFT-based architecture with low-complexity pre- and post-chirp operations. Owing to these advantages, organizations such as the University of Surrey, ETRI, and Shanghai Jiao Tong University have proposed AFDM as a complementary waveform to OFDM in 3GPP TSG-RAN meetings (e.g. R1-2604486, R1-2602803 and R1-2602781) \cite{R1_2600999}.

Despite the growing research interest, a systematic study of AFDM from a standardization perspective remains absent in the literature. This article addresses this gap by examining AFDM as a practical waveform candidate for future wireless networks. Rather than focusing solely on link-level error rate performance, we evaluate AFDM from a broader system perspective, encompassing backwards compatibility with legacy systems, support for multiple antennas and multi-user transmission, energy and spectral efficiency, implementation complexity, and emerging applications. \textit{In a nutshell, we investigate how AFDM can be seamlessly integrated into existing frameworks, what critical capabilities it offers for standardization, and where its most promising application opportunities lie.}

\section{AFDM for Fugure Wireless Networks}

\subsection{Preliminaries of AFDM}

Unlike conventional OFDM, which relies on   sinusoidal subcarriers,
AFDM is built upon chirp subcarriers whose instantaneous frequency varies
linearly with time.   For an AFDM symbol  of
$N$ subcarriers, the $n$th discrete-time chirp subcarrier is expressed as
 $
\psi_n[m] = \frac{1}{\sqrt{N}}\,
e^{j2\pi \left(c_1 n^2 + c_2 m^2 + \frac{nm}{N}\right)},
\quad m, n = 0, \ldots, N-1.$
Here, $c_1$  imposes a uniform chirp rate  across  
$N$ subcarriers and  $c_2$ introduces subcarrier-dependent initial
phases. Notably, OFDM and OCDM are both
special cases of AFDM, corresponding to $c_1=c_2=0$ and
$c_1=c_2=-1/(2N)$, respectively.   Therefore, as a natural
generalization of OFDM, AFDM  preserves strong backwards 
compatibility with legacy infrastructures. An important feature of the AFDM chirp subcarrier structure is spectral
wrapping upon digital sampling, as shown in Fig. \ref{AFDMTro}(a). Denoting the system bandwidth by $B$
and the symbol duration by $T = N/B$, each AFDM subcarrier is obtained
by sampling the continuous-time chirp signal  $K_n(t)  = \frac{1}{\sqrt{N}} {{\rm e}^{j 2 \pi  ({  {\frac{c_1 N^2}{ T^2}}t^{2}  +\frac{  n  t}{T}+ c_2n^2  })  }} $ at the sampling rate of $B$.   When $c_1 > 1/(2N)$, the instantaneous bandwidth of
$K_n(t)$ exceeds the Nyquist rate $B$, and digital sampling therefore
folds the spectrum periodically into the interval $[-0.5B,\, 0.5B]$, leading  to the wrapped subcarriers.
   A particularly
structured case arises when $c_1=Q/(2N)$ for some integer $Q$.
Under this condition, the $\frac{N}{2}$th subcarrier undergoes exactly
$Q$   spectral sweeps within one symbol duration, whereas the
remaining subcarriers complete exactly $Q+1$ sweeps\cite{Luo2026ChirpBasedOCDMAFDM,bemani2023affine}.

A key advantage of AFDM is its ability to exploit channel diversity over doubly dispersive channels. By properly selecting $c_1$, different delay-Doppler paths may occupy non-overlapping diagonals in the effective channel matrix, each path contributes a unique band-diagonal component in the DAFT domain. The resulting sparse, structured representation enables efficient suppression of both inter-symbol interference (ISI) and ICI even under severe delay and Doppler spreads, as shown in Fig. \ref{AFDMTro}(b).

\begin{figure*}
    \centering
    \includegraphics[width=0.8\linewidth]{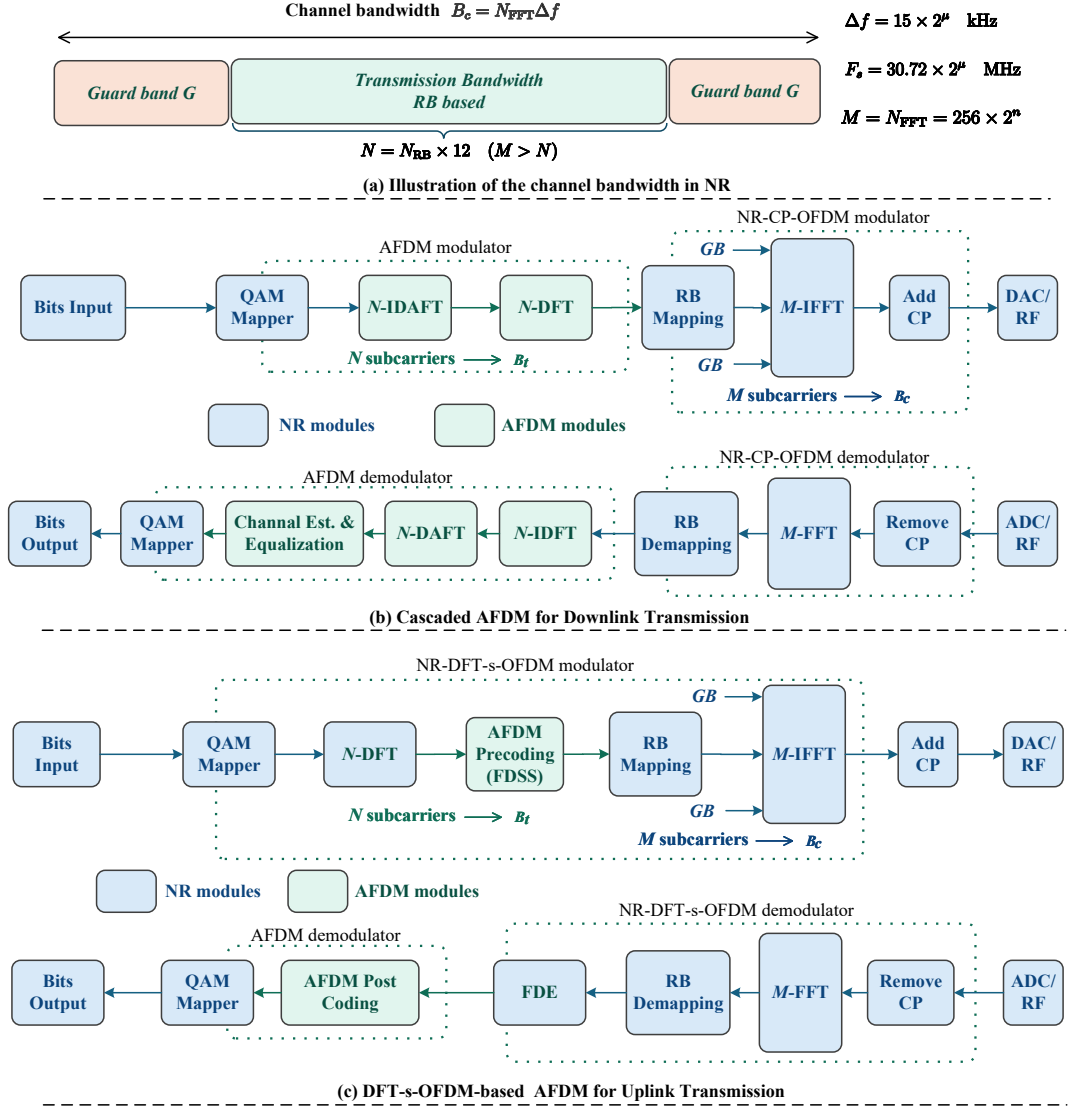}
    \caption{Downlink and Uplink backwards  Compatibility with 5G NR Numerology, where $c_1= \frac{Q}{2N}, Q \in \mathbb Z$ (DAC: digital-to-analog converter; CP: cyclic prefix; RF: radio frequency; QAM: quadrature amplitude modulation; GB: Guard band.).  }
    \label{AFDMDUPcom}
\end{figure*}

\subsection{    Standardization Considerations}
The aforementioned advantages make AFDM a highly promising waveform candidate for future wireless networks.  Standardizing AFDM, however,
requires a systematic evaluation that extends well beyond error-rate
performance. In general, waveform standardization involves a broad set
of considerations, including backwards and forwards compatibility, multi-antenna and multi-user support, spectral efficiency,
coverage,  peak-to-average power ratio (PAPR),   and scheduling flexibility.   Among these, backwards  compatibility is a major    consideration, since a practical waveform should integrate smoothly into the existing 5G   framework, including downlink  cyclic-prefix OFDM (CP-OFDM) and uplink  discrete Fourier transform-spread OFDM (DFT-s-OFDM), without radical  changes to the numerology, resource grid, or transceiver architecture.        
Meanwhile, future networks are expected to support emerging applications such as NTN, integrated sensing and communication (ISAC),  vehicle-to-everything (V2X), and underwater acoustic(UWA) communications, many of which generally operate under severe delay and Doppler dispersion.   In this regard, AFDM represents a timely and compelling opportunity.

\begin{figure*}
    \centering
    \includegraphics[width=1\linewidth]{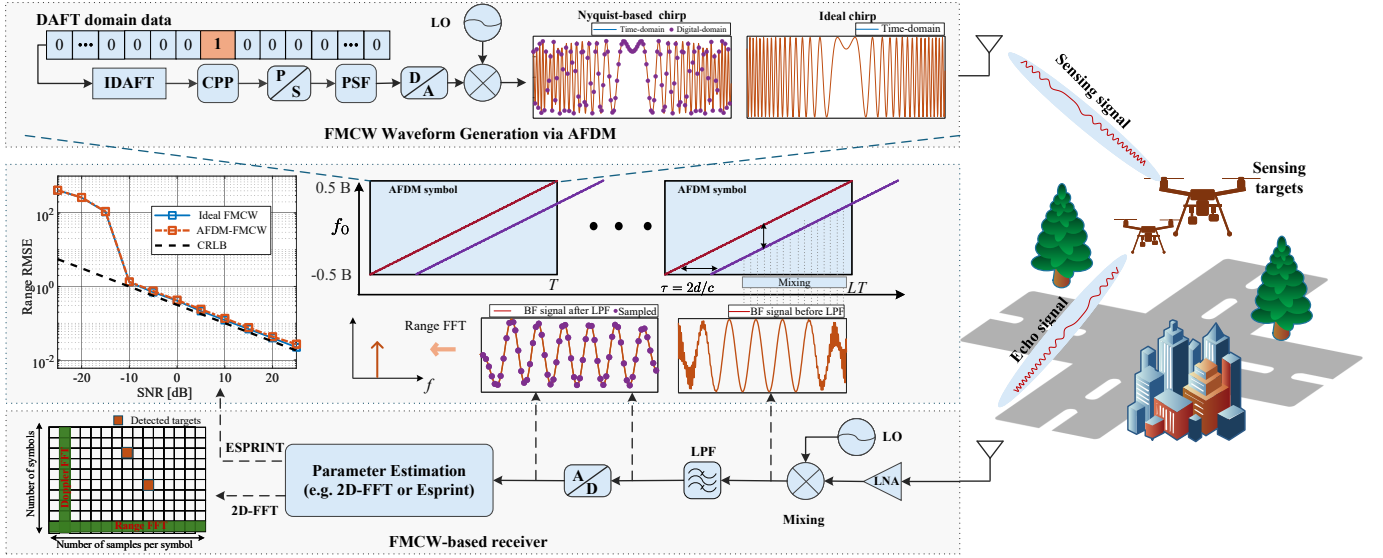}
    \caption{Illustration of the backwards  compatibility of AFDM with FMCW radar.}
    \label{AFDMFMCW}
\end{figure*}

 \section{backwards  Compatibility of AFDM }

\subsection{Downlink Backwards  Compatibility with   5G NR Numerology }

One of the defining features of 4G and 5G NR  is its flexible numerology, which enables scalable waveform parameterization for diverse deployment scenarios. Specifically, NR adopts subcarrier spacings of $\Delta f = 15 \times 2^{\mu}$ kHz, where $\mu \in \{0,1,2,3,4\}$ denotes the numerology index, as specified in TS 38.101 and TS 38.104 \cite{3GPP_38Series}, supporting subcarrier spacings up to $240$ kHz. To accommodate these numerologies within a unified framework, NR retains the legacy reference sampling rate of $30.72$ Msps, corresponding to a $2048$-point OFDM system with $15$ kHz subcarrier spacing. The basic sampling rate scales with numerology as $F_s = 30.72 \times 2^{\mu}\ \text{Msps}$.

Although the $N$-point IDAFT can be decomposed into an $N$-point IFFT kernel followed by low-complexity pre-chirp and post-chirp operations in the digital baseband, as shown in Fig. \ref{AFDMTro}(b), enabling backwards-compatible integration of AFDM into NR systems with multi-numerology operation requires addressing several practical issues. First, in 4G/NR downlink CP-OFDM transmission, the channel bandwidth is $B_c = B_t + G = N_{\text{FFT}}\Delta f$, as shown in Fig. \ref{AFDMDUPcom}(a). In OFDM, the guard band is realized by nulling edge subcarriers before the $N_{\text{FFT}}$-point IFFT. In AFDM, however, each DAFT-domain symbol spans the entire bandwidth, so nulling DAFT-domain symbols does not create guard bands in the same manner. Second, practical NR transmitters employ digital interpolation and filtering before the DAC to satisfy spectral emission requirements. Since AFDM inherently occupies the full bandwidth, interpolation-induced spectral aliasing is more difficult to suppress using conventional NR processing. Fig. \ref{AFDMDUPcom}(b) illustrates the   cascaded AFDM-NR waveform architecture for downlink transmissions. AFDM modulation is first applied over the transmission bandwidth $B_t$ via the IDAFT, followed by an additional FFT module and the conventional NR CP-OFDM processing chain. The resulting FFT-domain AFDM signal is mapped onto the active NR resource blocks (RBs)\footnote{An RB is the minimum scheduling unit in NR, consisting of $12$ subcarriers.}, while edge subcarriers are nulled to form the guard band. In this way, AFDM occupies a transmission bandwidth of $B_t$, while the subsequent CP-OFDM modulator determines the overall channel bandwidth. 


\subsection{Uplink backwards  Compatibility with  5G NR Numerology  }

In  the uplink, AFDM can be integrated in a backwards-compatible manner within the legacy DFT-s-OFDM architecture adopted by both 4G LTE and 5G NR.  In particular,  the chirp operation in the time domain can be equivalently realized as multiplicative spectral shaping in the frequency domain \cite{FinaChirp}.  AFDM can thus be embedded into DFT-s-OFDM by inserting a properly designed precoding matrix after DFT spreading \cite{SavauxDftafdm}, as shown in Fig. \ref{AFDMDUPcom}(c).   
This approach is highly compatible with existing 4G/5G uplink transmission. AFDM precoding can be realized through frequency-domain spectral shaping (FDSS), which has already standardized in 5G NR for PAPR reduction.  AFDM-oriented FDSS can thus be viewed as a structured extension of existing shaping methods, where the coefficients are designed not only for spectrum containment but also for realizing the target chirp-domain waveform. Since AFDM is incorporated as an FDSS-based processing block rather than a new modulation chain, the legacy uplink architecture, including DFT spreading, resource mapping, OFDM modulation, and CP insertion,  can be retained with minor modification. Moreover, AFDM naturally operates under the NR multi-numerology framework, reusing the same scalable FFT sizes, symbol durations, CP lengths, and sampling rates, enabling coexistence with legacy DFT-s-OFDM across different bandwidth parts and deployment scenarios.

In a nutshell, AFDM can be interpreted not as a replacement of the NR uplink framework, but as an enhanced signal-shaping mode embedded into the existing   transmitters. More importantly, AFDM can also serve as a forward-compatible waveform candidate for future standards, maintaining continuity across generations.

\subsection{ Backwards  Compatibility with Radar-centric  Waveform }

Chirp signals are well known for their \textit{pulse compression} capability and are therefore widely used in conventional radar systems, particularly  frequency-modulated continuous-wave (FMCW) radar \cite{xu2023automotive}.  In AFDM, when the $N/2$-th subcarrier is activated and  $c_1=\frac{1}{2N}$, the generated discrete chirp spans from $-0.5B$ to $0.5B$ over a chirp duration $T$ with chirp rate $\alpha= B/T$, which is compatible with the   discrete FMCW waveform, as illustrated in Fig. \ref{AFDMFMCW}.    At the receiver, the echo is mixed with a reference chirp to produce the beat-frequency (BF) signal at $f_{\text{BF}}=\alpha\tau$, where $\tau$ is the propagation delay.

Traditional FMCW radars operate in the analog domain via voltage-controlled oscillators, whereas AFDM generates chirps digitally through DAFT-domain modulation—a Nyquist-sampled discrete-time realization of FMCW \cite{Zhu2025ISACAFDMFMCW}. Although slight distortion may appear near band edges compared with ideal analog FMCW, these distortion components are shifted to high frequencies after dechirping (well above the desired BF term $\alpha\tau$) and are effectively removed by subsequent low-pass filtering (LPF), as shown in Fig.~\ref{AFDMFMCW}. Consequently, AFDM-FMCW achieves sensing performance close to conventional FMCW and near the Cramér--Rao lower bound (CRLB). This compatibility provides useful insights for future 6G ISAC standardization, particularly for V2X scenarios where FMCW radar is widely adopted.

\begin{figure}
    \centering
    \includegraphics[width=1\linewidth]{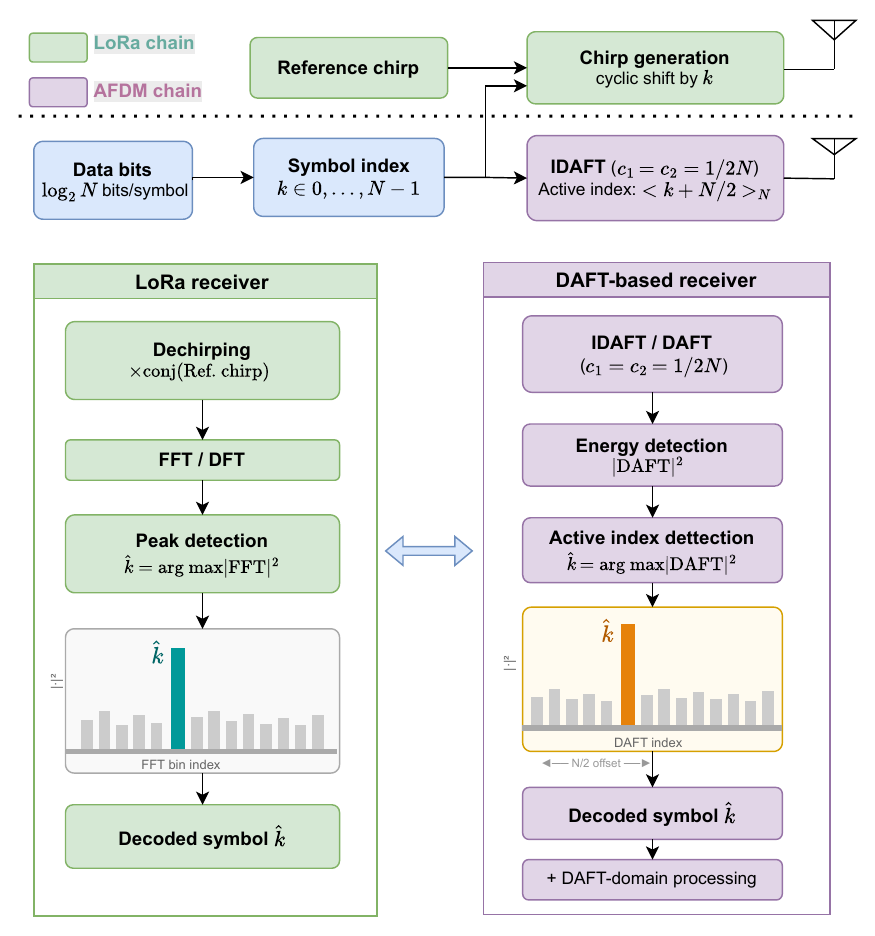}
    \caption{Illustration of the conventional LoRa modulation and the proposed AFDM-based LoRa tranceiver.}
    \label{LoraAFDM}
\end{figure}

\subsection{Backward Compatibility with LoRa Waveform} 
LoRa is a chirp spread spectrum (CSS) waveform widely used in low-power wide-area networks (LPWANs)  for Internet-of-Things (IoT)  applications. CSS encodes information through cyclic shifts of a reference linear chirp with each symbol corresponds to one of $N$ possible chirp states, carrying $\log_2 N$ bits. When AFDM parameters are set to $c_1=c_2=\frac{1}{2N}$, the active DAFT-domain index becomes exactly the data-carrying symbol position in conventional LoRa. Hence, LoRa signaling can be interpreted as a single-active-index chirp mode within AFDM, as shown in Fig. \ref{LoraAFDM}.

 This equivalence naturally yields transceiver backward compatibility, as shown in Fig. \ref{LoraAFDM}. The receiver can either reuse the conventional LoRa demodulation chain (dechirping followed by FFT-domain peak detection) or perform detection directly in the AFDM DAFT domain by identifying the active index position. The former preserves the existing LoRa receiver architecture, whereas the latter enables native AFDM-domain processing. This dual interpretation supports cross-domain processing between time and DAFT domains, offering enhanced robustness in high-mobility scenarios such as NTN LPWANs applications. From a standardization perspective, this connects AFDM to the existing LPWAN and narrowband IoT ecosystem.

\subsection{ Other Important Considerations  for Standardization  }

Beyond the compatibility aspects discussed above, waveform standardization also requires evaluation across multiple performance dimensions, including support to multiple-input and multiple output (MIMO) and multiple access (MA), and  energy efficiency.  Therefore, this subsection focuses on these three key capabilities of AFDM.

\textbf{1) Multiple access   }

MA capability  is an important consideration in waveform standardization.  Unlike the downlink where transmission is centrally coordinated at the base station, uplink access needs to accommodate independently transmitted user signals under heterogeneous channel conditions and asynchronous transmissions.     AFDM can support uplink MA along the following two complementary directions.

\textit{Native AFDM-oriented MA:} Similar to DFT-s-OFDM, a native DAFT-spread affine frequency division multiple access (DAFT-s-AFDMA) scheme was proposed in \cite{AFDMMADAFT}, where multiple users share the available chirp subcarriers and each user performs DAFT-domain spreading before resource mapping and AFDM modulation. By spreading each user's symbols over the allocated DAFT-domain resources, this structure is more robust to localized multi-user interference caused by heterogeneous delay-Doppler effects. Additionally, the DAFT spreading introduces a configurable pre-chirp parameter ($c_2$), providing an extra degree of freedom for PAPR reduction.

\textit{MA build upon DFT-s-OFDM:} Another attractive approach is to support AFDM-based uplink access through backwards-compatible integration into the existing 5G NR uplink framework. Since NR already adopts DFT-s-OFDM for uplink transmission, and AFDM can be naturally realized through FDSS-based DFT-s-OFDM processing, different users may employ user-specific FDSS coefficients to induce the desired chirp-domain structure, followed by DFT spreading and subcarrier mapping prior to OFDM modulation. This allows the existing scheduling and resource allocation mechanisms to remain largely unchanged, while naturally supporting coexistence of legacy DFT-s-OFDM users (e.g., low-mobility) and AFDM-enabled users (e.g., high-mobility) within a common uplink framework.

\begin{table}  
\footnotesize
    \caption{Achieved Net Gain of AFDM.}
    \centering
    \begin{tabular}{c|c|c}
    \hline\hline
    \textbf{UE Speed} & \textbf{Benchmark Waveform} & \textbf{Net Gain $@10\%$ BLER} \\
    \hline\hline
    \multirow{2}{*}{$3$ km/h} & CP-OFDM & $0.8$ dB \\
    \cline{2-3}
    & DFT-s-OFDM & $0.6$ dB \\
    \hline
    \multirow{2}{*}{$500$ km/h} & CP-OFDM & $3.0$ dB \\
    \cline{2-3}
    & DFT-s-OFDM & $2.5$ dB \\
    \hline
    \multirow{2}{*}{$1500$ km/h} & CP-OFDM & $5.2$ dB \\
    \cline{2-3}
    & DFT-s-OFDM & $4.5$ dB \\
    \hline
    \end{tabular}
    \label{sim_para}
    \\
    \raggedright
    \vspace{0.4em}
    Carrier frequency: $4$ GHz, $N = 512$, subcarrier spacing: $30$ kHz, channel model: TDL-A, modulation: QPSK, channel code: 5GNR LDPC with code length and rate of $512$ and  $ 4/5$, respectively. 
    \vspace{-2em}
\end{table}

\textbf{2)  PAPR  }

PAPR is a key concern in waveform standardization, particularly for uplink transmission due to UE power amplifier constraints. In recent 3GPP discussions on low-PAPR uplink waveforms, the evaluation focuses on the \emph{Net Gain} criterion under the same spectral efficiency as the reference waveform:
\begin{equation}
G_{\mathrm{net}} = G_{\mathrm{tx}} - \Delta\gamma_{10\%\mathrm{BLER}},
\label{eq:netgain}
\end{equation}
where $G_{\mathrm{tx}}$ denotes the transmit-power gain relative to the reference waveform, and $\Delta\gamma_{10\%\mathrm{BLER}}$ denotes the SNR degradation at $10\%$ block error rate (BLER). Under this model, Net Gain thus reflects the   transmit-power gain   jointly with the BLER-related SNR loss,  rather than a standalone PAPR reduction. A realistic PA model and PA back-off should also be incorporated when evaluating such gains.  

 AFDM is attractive in that its PAPR reduction can be achieved along the following two    directions.  First, both native AFDM and its DFT-s-OFDM-based realization remain compatible with existing low-PAPR techniques (PTS, tone reservation, etc.). Second, the chirp parameter $c_2$ can generate different DAFT-domain phase rotations, leading to signals with different peak characteristics—analogous to selected mapping (SLM) in OFDM but induced by intrinsic chirp parameterization \cite{yuan2024papr,AFDMMADAFT}.  Table \ref{sim_para} compares the Net Gain of AFDM over CP-OFDM and of DAFT-s-AFDM over DFT-s-OFDM, respectively, using a memory-based polynomial PA model. As observed, the Net Gain of AFDM over OFDM becomes more prominent as mobility increases.  

  \textbf{3) MIMO}
  
MIMO-AFDM systems process signals in a joint spatial-Affine (SAF) domain, combining AFDM's delay-Doppler resilience with MIMO spatial gains. The end-to-end relation in the SAF domain is $\mathbf{y} = \sum_{i=1}^{P} (\mathbf{H}_{\text{S},i} \otimes \mathbf{H}_{\text{AF},i})\mathbf{Q}\mathbf{x} + \mathbf{w}$, where the per-path SAF channel is the Kronecker product of spatial ($\mathbf{H}_{\text{S},i}$) and affine-domain ($\mathbf{H}_{\text{AF},i}$) channel matrices \cite{mimoafdmstad}. Since spatial and affine domains capture inherently different physical characteristics, precoding must be designed jointly across both. In single-user scenarios, the Kronecker structure enables a low-complexity LOS-based approximation requiring only LOS-path CSI at the transmitter \cite{mimoafdmstad}. In multi-user settings, weighted MMSE extends naturally to the SAF domain to suppress inter-user interference across spatial, delay, and Doppler dimensions jointly.

\section{Standardizing AFDM Towards Emerging Applications } 
 AFDM is particularly attractive in scenarios where robustness to delay--Doppler dispersion is  of central importance. In this regard, AFDM exhibits strong potential in several emerging applications, including NTN, ISAC, V2X, and UWA communications. This sections discuss how the intrinsic properties of AFDM can be exploited to meet/support these requirements and evolution trends of these emerging application domains.

\subsection{NTN}
\begin{figure}[!t]
    \centering
    \subfloat[$30$ km/h, TDL-A]{
        \includegraphics[width=0.45\linewidth]{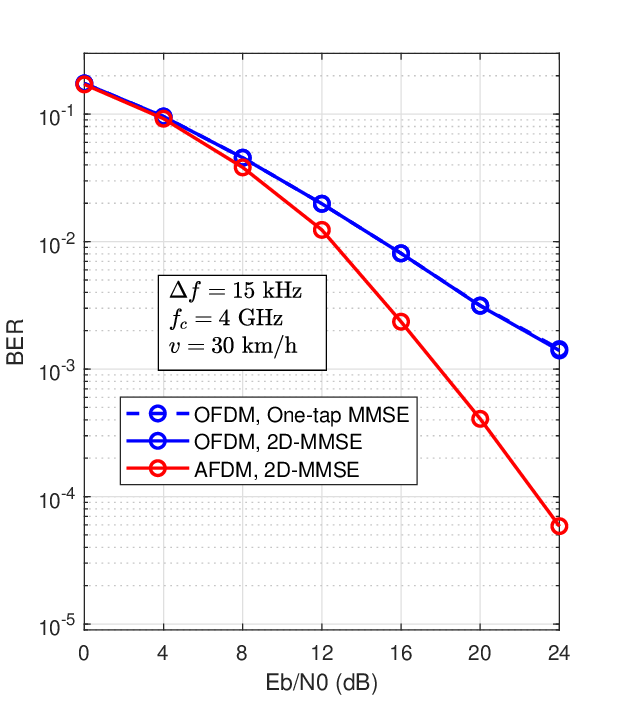}
        \label{fig:sub1}
    }
    \hfill
    \subfloat[$500$ km/h, TDL-A]{
        \includegraphics[width=0.45\linewidth]{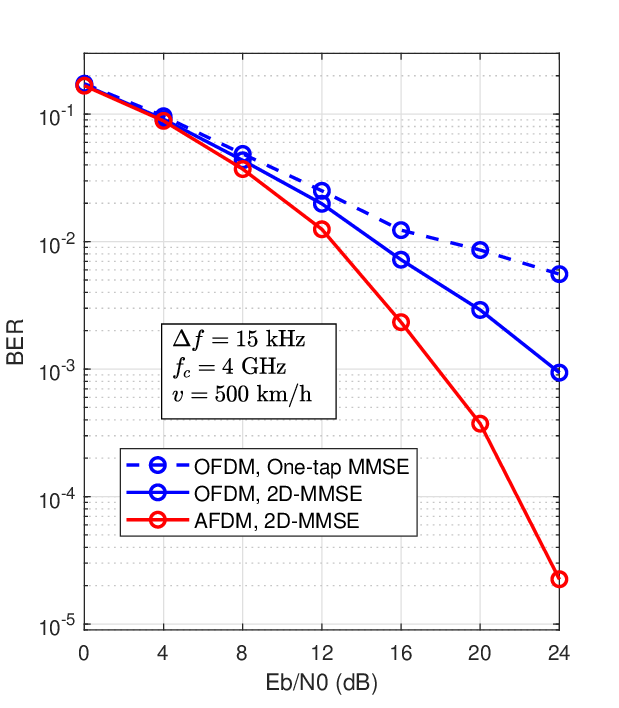}
        \label{fig:sub2}
    }

    \vspace{0.3cm}

    \subfloat[$500$ km/h, TDL-D]{
        \includegraphics[width=0.45\linewidth]{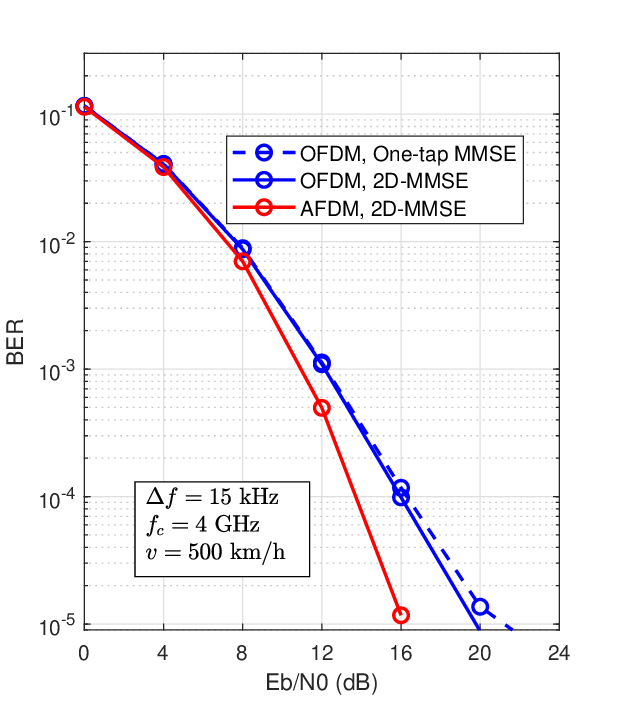}
        \label{fig:sub3}
    }
    \hfill
    \subfloat[$500$ km/h, TDL-D]{
        \includegraphics[width=0.45\linewidth]{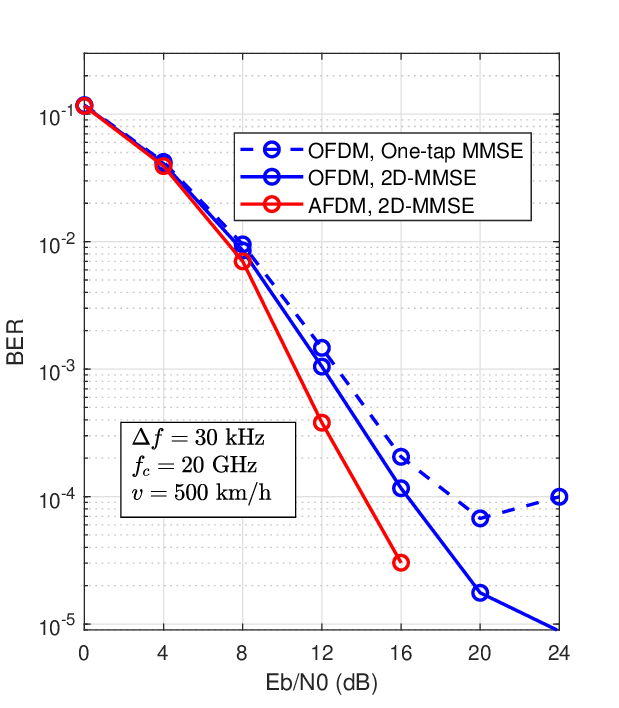}
        \label{fig:sub4}
    }
    \caption{BER comparisons of OFDM and AFDM over NTN-TDL channels. (\small{The delay and power profiles of the first three taps for NTN-TDL-A are $\boldsymbol{\tau}_{\text{A}} = [0,\; 38.2,\; 40.3]$~ns and $\mathbf{p}_{\text{A}} = [-13.4,\; 0,\; -2.2]$~dB, and those for NTN-TDL-D are $\boldsymbol{\tau}_{\text{D}} = [0,\; 3.5,\; 61.2]$~ns and $\mathbf{p}_{\text{D}} = [-0.2,\; -18.8,\; -21.0]$~dB, respectively. The actual tap delays are obtained as $\tau_l = \tau_{l,\mathrm{norm}} \times \mathrm{DS}$, where $\mathrm{DS}$ denotes the  root-mean-square delay spread, which is set to $100$~ns.  The complete model parameters can be found in~\cite{3GPP_38Series}}).
     } 
    \label{ntnber}
\end{figure}

NTN  are a key enabler for ubiquitous connectivity in next-generation wireless networks. Typical NTN platforms include UAVs, high-altitude platforms, aircraft, and satellites, all inherently associated with high-mobility environments. In satellite communications, NTN links are further characterized by large propagation delays and delay variations. To cope with Doppler effects, legacy OFDM-based solutions mainly rely on two mechanisms: 1) increasing the subcarrier spacing  at the cost of shorter symbol duration and reduced robustness to long delay spread; and 2) estimating and compensating the dominant Doppler component while treating residual Doppler as interference. However, this strategy becomes increasingly ineffective in downlink NTN scenarios where multipath propagation is prevalent and different paths may experience distinct delay-Doppler shifts, leading to non-negligible residual interference.   Fig. \ref{ntnber} compares the BER of AFDM and OFDM over the 3GPP NTN-TDL channel models specified in TR 38.901 and TR 38.821 \cite{3GPP_38Series}, where NTN-TDL-A/B corresponds to non-LOS scenarios and NTN-TDL-C/D to LOS-dominant conditions. The ``OFDM, One-tap'' curves employ conventional one-tap linear MMSE equalization with dominant-path Doppler compensation. In LOS-dominant scenarios, OFDM remains applicable when the dominant Doppler is accurately compensated.   However, in richly scattered  channels where different paths experience non-negligible delay-Doppler coupling, conventional Doppler compensation in OFDM may be  ineffective.

Additionally, the initial access procedure in high-mobility satellite networks  itself is  a major challenge for OFDM-based systems. Random-access preambles must be detected under large delay uncertainty, strong Doppler shift, and Doppler drift. Current NR NTN applies GNSS-based Doppler pre-compensation and Zadoff-Chu preamble sequences up to length 839 for roubust initial access. Nevertheless, reliable preamble detection remains challenging under large delay uncertainty, severe Doppler shift, and Doppler drift.  By contrast, AFDM provides additional flexibility: its chirp waveform preserves favorable correlation characteristics, and the DAFT-domain representation facilitates separation of Doppler shifts, multipath delays, and their coupled dispersion, making it attractive for robust preamble detection in satellite scenarios.


 \subsection{ISAC }

\begin{figure*}
    \centering
    \subfloat[A comparison of communication- and sensing-centric AFDM-ISAC systems.]
        {\includegraphics[width=0.9\linewidth]{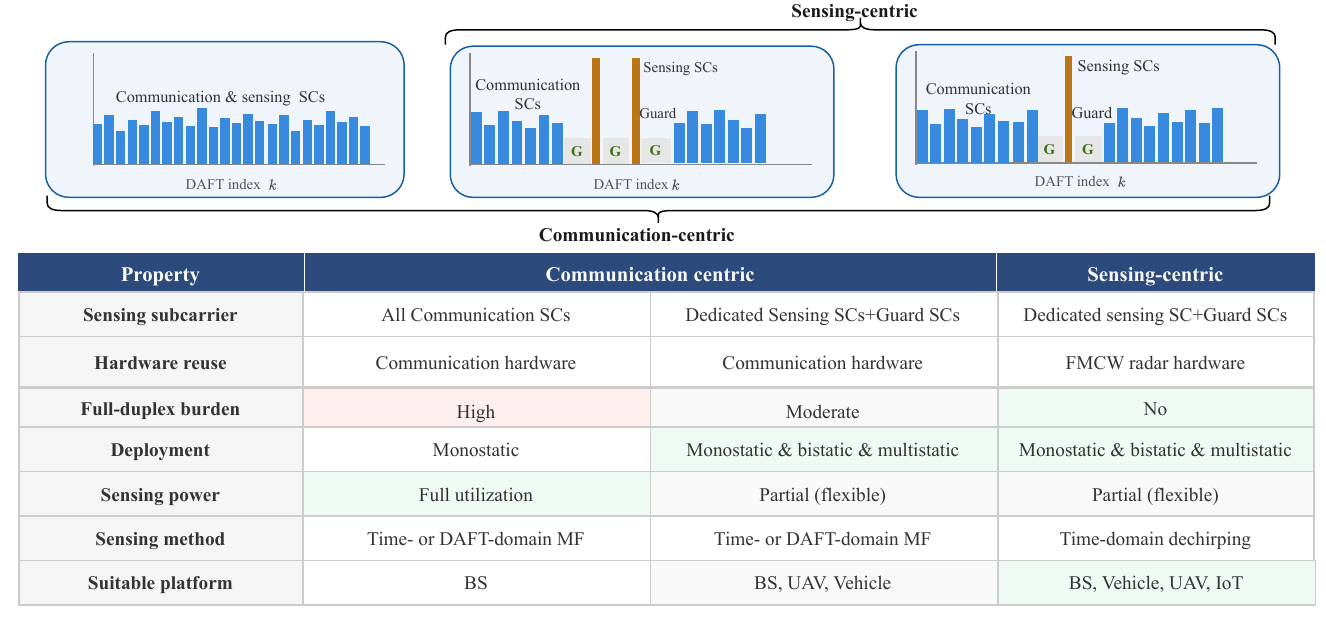}\label{fig:ISAC1}}
    \\[6pt]
    \subfloat[Illustration of sensing-centric AFDM-ISAC design.]
        {\includegraphics[width=0.9\linewidth]{figures/ISAC2.pdf}\label{fig:ISAC2}}
    \caption{Illustration of communication- and sensing-centric AFDM-ISAC systems.}
    \label{fig:ISAC}
\end{figure*}
ISAC is widely recognized as an important capability for future wireless standards. Owing to its time-domain chirp pulse compression and DAFT-domain representation, AFDM provides several attractive advantages for ISAC applications \cite{11164334}. In particular, AFDM supports both communication-centric ISAC, where communication signals are directly reused for sensing with maximum reuse of existing hardware, and sensing-centric ISAC, where deterministic chirp signaling is employed for radar sensing with compatibility to existing FMCW radar architectures. Fig. \ref{fig:ISAC} summarizes the distinct features of these two schemes and provides a comprehensive comparison.

\textbf{Communication-centric ISAC:}

AFDM aided  communication-centric sensing can be realized in the following two  modes.

\textit{Full-subcarrier sensing:} All AFDM chirp subcarriers simultaneously carry communication data and contribute to sensing. Target delay and Doppler are estimated via time-domain or DAFT-domain matched filtering after transmitted-data cancellation. This mode fully exploits the transmitted signal energy, thereby maximizing sensing SNR, and the DAFT-domain processing integrates naturally into the communication receiver chain. However, since sensing relies on knowledge of the transmitted data, this mode is mainly suitable for monostatic sensing with co-located transmitter and receiver, and requires effective full-duplex operation with sufficient antenna isolation to prevent ADC saturation caused by RF transmit-signal leakage.

 \textit{Partial-subcarrier sensing:} Since each AFDM chirp subcarrier occupies the entire bandwidth in time, sensing can be realized using only one or a few deterministic subcarriers while the remainder serve guard and communication purposes. This mode offers much greater flexibility in communication-sensing resource allocation, e.g., the sensing power can be adjusted according to requirements. It also alleviates the strict full-duplex requirement, as digital self-interference cancellation (SIC) can be implemented directly in the DAFT domain. Moreover, deterministic sensing subcarriers are promising for bistatic and multistatic scenarios, where they simplify waveform coordination and improve the scalability of distributed sensing.

\textbf{Sensing-centric ISAC:}

In sensing-centric ISAC, dedicated chirp subcarriers (i.e., partial subcarriers) are designed primarily for sensing, while existing radar signal processing techniques and hardware architectures can be largely reused. As discussed in Subsection  III-C, when AFDM is properly configured, the receiver can directly reuse the  existing FMCW radar processing  architecture. Fig. \ref{fig:ISAC}(b) shows the proposed AFDM-ISAC by using the $N/2$th subcarrier for sensing  with $c_1=1/N$.  Dechirping between communication subcarrier echoes and the reference signal also generates BF components. However, by introducing a proper guard band, the BF components from communication subcarriers can be effectively suppressed through LPF design.  For larger chirp rates, i.e., $c_1=Q/(2N), Q>1$, , spectral wrapping occurs for AFDM chirp subcarriers in the frequency domain. This spectral wrapping, together with CPP insertion, introduces high-frequency components and phase discontinuities in the dechirped BF signal across the delay and Doppler dimensions. To address this issue while preserving compatibility with existing FMCW radar receivers, we propose two AFDM-oriented chirp-fusion schemes, as shown in Fig. \ref{fig:ISAC}(b). This dechirping-based architecture is particularly attractive as it avoids   full-duplex operations, suppresses local leakage and communication-induced interference, and enables low-complexity implementation based on existing FMCW hardware.

Overall, AFDM    provides a distinctive ISAC standardization opportunity by supporting both communication-centric and sensing-centric modes within a unified chirp-based framework. For communication-centric ISAC, AFDM enables either full-band sensing to maximize sensing SNR or sparse sensing-subcarrier operation for improved flexibility and reduced full-duplex burden. For sensing-centric ISAC, AFDM offers strong compatibility with FMCW-based dechirping receivers, leading to hardware-efficient radar-oriented designs.
 
\subsection{V2X}
V2X systems demand low-latency, high-reliability information exchange among vehicles, infrastructure, pedestrians, and networks. AFDM's chirp-based structure and robustness in doubly dispersive channels make it well suited for synchronization signaling, random-access preamble design, and broadcast safety-message transmission under high-mobility V2X systems, aligning with the evolution of NR V2X sidelink and future vehicular PHY. Additionally, AFDM's ISAC capability supports the growing need for environment sensing and positioning in vehicular networks. From a standardization perspective, AFDM offers a promising enhancement for extreme-mobility sidelink, robust broadcast/control signaling, and future ISAC-enabled vehicular transmission.
\subsection{UWA}
UWA is another representative scenario in which AFDM may offer distinct advantages. UWA channels exhibit limited bandwidth, long propagation delay, large delay spread, severe Doppler distortion, and difficult synchronization, making reliable waveform design far more challenging than in terrestrial systems. AFDM is particularly suitable as a robust waveform for UWA control, discovery, and synchronization signaling. Its DAFT-domain structure is well matched to channels with strong multipath and non-negligible Doppler effects, improving waveform resilience in underwater links. In this context, AFDM is positioned not as a throughput-maximizing waveform, but as a robust PHY framework for foundational signaling and dependable low-rate communication under harsh underwater conditions.

 \section{Conclusion and  Future Directions}

This article has presented a systematic study of AFDM from a standardization perspective for future wireless networks. We demonstrated that AFDM exhibits strong backwards  compatibility with the existing 5G NR downlink and uplink frameworks, as well as with FMCW radar waveforms, LoRa modulations, enabling its integration with minimal modifications to legacy processing chains. Key standardization-critical capabilities, including multiple access, PAPR reduction, and MIMO support, were evaluated, showing that AFDM provides both waveform-native design flexibility and compatibility with existing NR mechanisms. Furthermore, AFDM demonstrates strong potential in emerging scenarios, including NTN, ISAC, V2X, and UWA communications, where severe delay-Doppler dispersion places stringent demands on waveform robustness. These features position AFDM as a timely and compelling waveform candidate for future wireless standards.

Several open challenges remain for AFDM standardization.  These including   chirp-parameter configuration and reference-signal design compatible with existing NR mechanisms, initial access and ISAC waveform design for high-mobility scenarios (e.g., V2X, NTN, UWA), waveform coexistence with legacy systems, scalable MIMO and multi-user support, and hardware-oriented validation for practical deployment.


\ifCLASSOPTIONcaptionsoff
  \newpage
\fi

\bibliographystyle{IEEEtran}
\bibliography{ref.bib}
 
\end{document}